\documentclass[%
utf8,
preprint,superscriptaddress,
amsmath,amssymb,prd,aps,showpacs
floatfix,
]{revtex4-1}
\usepackage{graphicx}
\usepackage{dcolumn}
\usepackage{bm}
\usepackage{mathrsfs}
\usepackage{amsmath}
\usepackage{amssymb}
\usepackage{float}
\usepackage{dcolumn}
\usepackage{multirow}

\begin{document}


\title{The path integral formula for the stochastic evolutionary game dynamics in the Moran process}
\author{Chao Wang}
\email[E-mail: ]{chaowang@nwpu.edu.cn}
\affiliation{School of Ecology and Environment, Northwestern Polytechnical University, Xi’an 710072, China}


\begin{abstract}
The Moran process is one of an basic mathematical structure in the evolutionary game theory. In this work, we introduce the formulation of the path integral approach for evolutionary game theory based on the Moran process. We derive the transition probability
by the path integral from the initial state to the final state with updating rule of the Moran process. In this framework, the transition probability is the sum of all the evolutionary paths. The path integral formula of the transition probability maybe
expected to be a new mathematical tool to explore the stochastic game evolutionary dynamics.
\end{abstract}

\maketitle


\section{Introduction}

The problem of explaining the cooperation behavior has been exploring for a long time. Since the 1970s, evolutionary game theory has achieved great success in explaining the evolution of animal behavior and been widely used in economics, social sciences and
human behavioral ecology, etc \cite{hofbauer1998,nash1950,smith1982,lawlor1976}. Traditional, the dynamic of the systems was studied by the well-know replicator equation \cite{Taylor1978}. Recently, the stochastic evolutionary game have been study in the finite population \cite{Nowak2004,Imhof2006} and within the impact of environmental stochastically \cite{Braga2022,Frank2022,Michael2013}. Several descriptions of the stochastic evolutionary game dynamics has been preformed, such as the Langevin equations and the stochastically equivalent Fokker planck equation \cite{traulsen2005}. However, all of those approaches have concentrated on the stationary distribution stable states (or the relative structure of stable states) and the information of the intermediate process is locked.

The path integral method, which was raised from Wiener \cite{1921The} for Brownian motion, has found substantial success in applications in quantum mechanics \cite{1948Space,1965Quantum} and quantum field theory \cite{Anthony2010Quantum}. In the 1940s Feynman
\cite{1948Space} formulated a space-time approach to quantum mechanics based on the path integral introduced by himself. Recent years have witnessed a majority efforts to develop a similar path integral formulation for the theory of the stochastic process \cite{kleinert2009}. And such a formulation can supply a calculational technique that may be fruitful in dealing with nonlinear systems. Although general evaluation of the path integral is impossible for nonlinear systems, its definition as the limit of a multidimensional integral suggests powerful numerical methods \cite{1983Numerical,M1983Numerical,1987Numerical}. Recently, Joshua has preformed attempting for the evolutionary problem of the gene frequency in the population genetics, showing how it is possible to use path integral methods to express the transition probability by means of a functional integral over paths of the evolutionary process for the Wright-Fisher process \cite{Schraiber2014}. The aim of this work is to provide a contribution to the evolutionary problem based on evolutionary game theory, showing how it is possible to use path integral methods to express the transition probability by means of a functional integral over paths of the evolutionary process.

The Moran process is a basic model for stochastic dynamics in finite population in evolutionary game theory \cite{Moran1962}. The Moran process allows one to derive the fixation probability or the fixation probability In the previous papers, the Moran process have been used to study the frequency dependent approach of evolutionary game theory \cite{Lieberman2005,Nowak2004}. In the present work, we will calculate the transition probability in the formulation of path integral by recurring one-step process from the view point of the Moran model directly, rather than the Fokker-Plank equation. The basic idea is dividing the time interval into small slices of a infinitesimal duration, and deriving the propagator in the small slices directly by the Moran update rule. After that, we associate each path from the initial point to the final point with a probability, and then integrate over all possible paths to find the transition probability between these two points.

This article is organized as follows. In Sect.~II, we give a description of the Moran model. In Sect.~III, we illustrate the framework of the path integral and introduce the transition probability of an evolutionary path.  In Sect.~IV  we derive the short-time
propagator of the Moran model with the selection intensity. The transition probability of a particular evolutionary path based on the Moran model and the corresponding path integral formula is then calculated. Some discussions and conclusion are given in the
last Sect.~V.

\section{the Moran process}
We start by considering a two-strategy stochastic evolutionary game by the Moran undate rule. The fitness (or payoff) of two strategies $A$ and $B$ depends on the composition of the population and is given by the payoff matrix
\begin{eqnarray}
  \begin{array}{ccc}
      & A & B \\
    A & a & b \\
    B & c & d \\
  \end{array}.
\end{eqnarray}
In an finite size ($N$) and well-mixing population, the expected payoff is $\frac{a(j-1)+b(N-j)}{N-1}$ for $A$-type and $\frac{cj+d(N-j-1)}{N-1}$ for $B$-type. As Nowak et, we take an selection intensity parameter $\omega$ ($0\leq\omega\leq1$) and the expected
payoffs are therefore given by
\begin{eqnarray}
f_j&=&(1-\omega)+\omega\frac{a(j-1)+b(N-j)}{N-1},\nonumber\\[4pt]
g_j&=&(1-\omega)+\omega\frac{cj+d(N-j-1)}{N-1}.
\end{eqnarray}
For large $N$, the fitness can be approximatively denoted by
\begin{eqnarray}
f_j&=&(1-\omega)+\omega[ax+b(1-x)],\nonumber\\[4pt]
g_j&=&(1-\omega)+\omega[cx+d(1-x)],
\end{eqnarray}
with $x=j/N$ being the frequency of the strategy $A$. For the Moran process, the corresponding transition probabilities from $x$ to $x+1/N$ and from $x$ to $x-1/N$ for each time step are then obtained:
\begin{eqnarray}\label{pi}
\pi^+&=&\frac{x(1-x)f(x)}{xf(x)+(1-x)g(x)},\nonumber\\[4pt]
\pi^-&=&\frac{x(1-x)g(x)}{xf(x)+(1-x)g(x)},
\end{eqnarray}
respectively.

It's reasonable to assume that the changing in strategy frequencies is Markovian, that is, the probability distribution of the frequency in the current generation depends on the gene frequency at the last generation but not on the previous history which has led
to the frequency at the last generation. Two well-known descriptions of such processes in the study of the their dynamics exist: the Langevin equations (and their mathematically refined versions: the Ito equation), and the stochastically equivalent Fokker planck
equation \cite{zheng2011diffusion,tran2013introduction}. They both allow to determine the stochastic properties of the system at the current time, if those properties are known at some earlier time. 

In the present work, we will concentrate on another parameter--the transition probability from one state to another state considering all the intermediate states it passed--in terms of path integral of such processes is derived in the following section. In that
description, the dynamics then can be formulated by writing down a probability density for observing a complete path of the system step by step \cite{1965Quantum}. Specifically, the transition probability can be seen as the convolution of an infinite sequence of
infinitesimal short time steps. We here will give the evolutionary dynamics in the formula of path integral.

\section{The transition probability}
\begin{figure}[ht]
  \centering
  \includegraphics[width=0.75\textwidth, keepaspectratio]{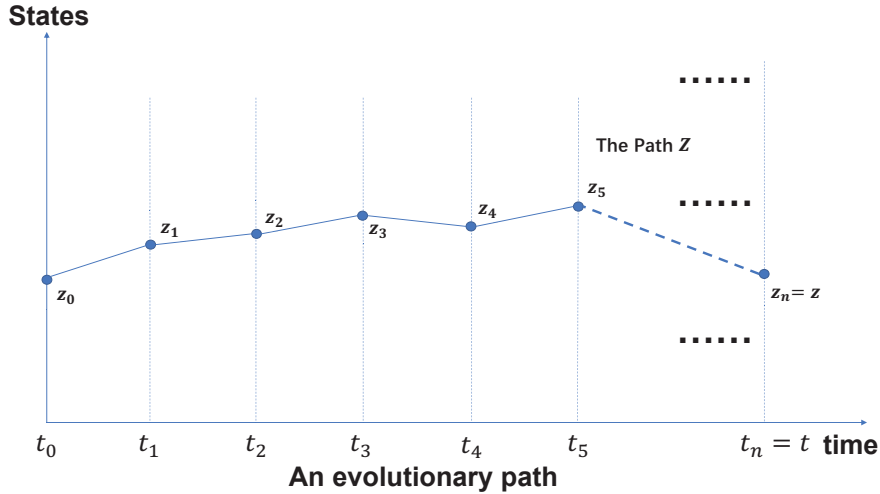}
  \caption{The evolutionary path. The time scale is divided into $N$ intervals, i.e. $\delta t= (t-t_0)/N$. For the Moran process which depicts the evolutionary process, $\delta t$ represents one generation. A way to summing up all the possible paths like this
  is followed. The transition probability from $(z_0,t_0)$ to $(z,t)$ is to do the sum over all possible states and time from $(z_0,t_0)$ to $(z_1,t_1)$, then from $(z_1,t_1)$ to $(z_2,t_1)$, and so on, until to $(z_{n-1},t_{n-1})$.}\label{fig1}
\end{figure}
\subsection{The framework of the path integral}
 Given an initial probability distribution, the statistics of the diffusion process can be completely calculated in terms of its transition probability \cite{1992Path}. The transition probability, denoted as $p(z,t\mid z_0,t_0)$, is defined as the probability of
 a transition from one point in the system state space, $z_0$, at the initial time $t_0$ to another point in the state space, $z$, at the final time $t$, where $t>t_0$. In such a transition, all the possible paths connecting these two points play roles, each of
 them realizing a sample function and therefore having a definite possibility. Then it is natural to postulate that the probability of the overall transition is a sum of all intermediate possibilities from $z_0$ to $z$ \cite{1965Quantum}:
\begin{equation}\label{eq5}
  p(z,t\mid z_0,t_0)=\sum_{all\ paths} \phi[Z(t)],
\end{equation}
where $\phi [Z(t)]$ is the contribution from a particular path $Z(t)$ (as shown in Fig. \ref{fig1}). and ``all paths" means of course all possible paths satisfying $z(t_0)=z_0$ and $z(t)=z$. It turns out that the ``sum" in Eq. (\ref{eq5}) can be realized as a
functional integral over the space of all possible paths (hence the name ``path integral"), with a probability distribution on the path space as its integrand.

\subsection{The transition probability of an evolutionary path}
\begin{figure}[ht]
  \centering
  \includegraphics[width=0.7\textwidth, keepaspectratio]{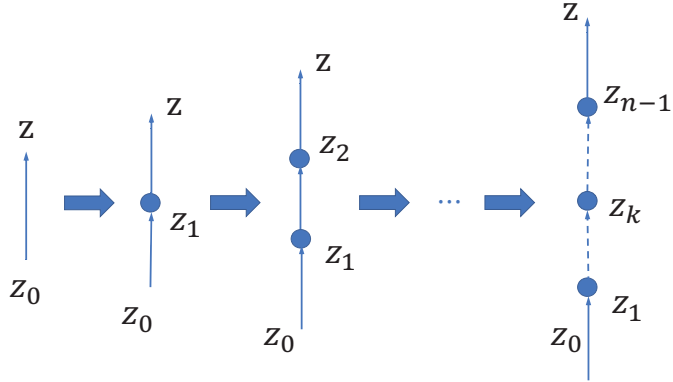}\\
  \caption{The division is made from the initial state to the final state for many times. This figure from the left side to the right side represents that the division is made by zero, once, twice, and $k$ times, respectively. Only if the division is much enough
  so that the short-time propagator for the stochastic process can be derived, the evolutionary path for the transition probability then can be derived.}\label{fig3}
\end{figure}

In this section we shall derive the calculation method for the transition probability for events which occur successively in time. As shown in Fig. \ref{fig3}, the initial point is $(z_0,t_0)$, and the final point is $(z,t)$. It's possible to find the system
state, $z_1$, at some time between the time $t_0$ and $t$. Then the transition probability along any path between $(z_0,t_0)$ and $(z,t)$ can be written. All alternative paths for the allele frequency from $z_0$ to $z$ can be labeled by specifying the position
$z_1$ through which the allele frequency passes at time $t_1$. Then the transition probability for the system state going from $z_0$ to $z$ can be computed. The transition probability to go from $z_0$ to $z$ is the sum, over all possible values of $z_1$, of the
transition probability to go from $z_0$ to $z_1$ and then from $z_1$ to $z$. Thus, the transition probability from $z_0$ to $z$ through $z_1$ can be writing as
\begin{equation}
P(\{z,\,z_1,\,z_0\})=p(z,t|z_1,t_1)p(z_1,t_1|z_0,t_0),
\end{equation}
where the symbol $\{\cdots\}$ means a evolutionary path $z_0,\,z_1,\,\cdots,\,z$. Similarly, it's possible to make two divisions in all the paths: one at $t_1$, and the other at $t_2$. That's because it's easy for biologists to obtain the timely frequency $z_2$
at the time $t_2$ between the time $t_1$ and $t$. Then the transition probability for the allele frequency going from $z_0$ to $z$ can be written
\begin{equation}
P(\{z,\,z_2,\,z_1,\,z_0\})=p(z,\,t|z_2,t_2)p(z_2,t_2|z_1,t_1)p(z_1,t_1|z_0,t_0).
\end{equation}
This means that we look at the frequency which goes from $z_0$ to $z$ as if it went first from $z_0$ to $z_1$, then from $z_1$ to $z_2$, and finally from $z_2$ to $z$. The transition probability taken over all such paths that go from $z_0$ to $z$ is obtained by
integrating this product over all possible values of $z_1$ and $z_2$.

We can continue this process until the time interval is divided into $N$ intervals (as shown in Fig. \ref{fig1}), i.e. $\delta t= (t^{'}-t_0)/N$. Let the state, which might result from measurement of the coordinate at time $t_{k}$, be $z_{k}$ (specifically,
$z_n=z, t_n=t$).
\begin{align}
P(\{z\}) & = p(z,t|z_{n-1},t_{n-1})p(z_{n-1},t_{n-1}|z_{n-2},t_{n-2})\times\cdots \notag\\[2pt]
  &\qquad\qquad \cdots \times p(z_2,t_2|z_1,t_1)p(z_1,t_{1}|z_0,t_{0}),
\end{align}
According to the conditional probability of stochastic process, the contribution from a particular path $\{z\}$ has the measure
\begin{align}
\phi[{z}]=P(\{z\})\Delta\{z\}
\end{align}
The probability of the overall transition is therefore
\begin{align}
 p(z,t\mid z_0,t_0)=\sum\limits_{\{z\}} P(\{z\})\Delta\{z\}
\end{align}

Eventually, we expect to go the limit $N\rightarrow \infty$, where the nodes of the paths is continuously infinite, and the limit $\Delta z \rightarrow \mathrm{d} z $, the transition probability is
\begin{align}
 p(z,\,t\mid x_0,\,t_0)  & =\int_{R}p(z,t|z_{n-1},t_{n-1})p(z_{n-1},t_{n-1}|z_{n-2},t_{n-2})\cdots\times\nonumber\\[2pt]
 &\qquad\qquad\times p(z_2,t_2|z_1,t_1)p(z_{1},t_{1}|z,t_{0}) \mathscr{D}z,\notag\\[4pt]
  & =\int_{R}\prod_{k=1}^{n}p(z_{k},t_{k}|z_{k-1},t_{k-1})\mathscr{D}z,
\end{align}
that the integration is to be taken over those ranges of the variables which lie within the region $R$, i.e., all continuous paths with constraints of $z_0$ ($t=t_0$) and $z$ ($t=t^{'}$). As is graphically depicted in Fig. \ref{fig3}, division is done once,
twice, ..., until $N$ times between the initial state and the final state. For simplicity, we denote $\mathscr{D}z=\mathrm dz_{1} \mathrm dz_{2}\ldots \mathrm dz_{n-1}$. $p(z_k,t_k|z_{k-1},t_{k-1})$ is the transition probability during a very short time, which
is so called short-time propagator. For the evolutionary game this propagator is calculated by one-step update process, since the infinitesimal step could be approximate to one update.

\section{the formulation of path integral for the Moran process}
Based on the concept of path integral mentioned above, we will explore the integrand in the evolutionary system whose evolutionary way is described by recurring the Moran process. We take the frequency distribution of the strategy $A$, $x_{t_i}$, at the time
$t_i$ as the system states. The probability of a path is a function of a series values of the state valuable $x_{t_0},\,x_{t_1},...,\,x_{t_{n-1}},\,x_{t_n}$. The probability of that path lies in a particular region $R$ of space-time thus is obtained by
integrating the integrand over the region. For the biological evolutionary system whose evolutionary way is the Moran process, random drift and selection intensity together lead to the various gene frequencies at each time, the range of which is $[0,\,1]$.

For the Moran process, the transition probability in a short time could been derived in Eq.~(\ref{pi}), which is the short-time propagator thought as the critical quantity in path integral approach. In the evolutionary process, there are two parts in the
transition probability when we consider the effect of selection intensity on the evolutionary dynamics. The former refers to the transition probability including natural selection and the latter refers to the situation which is completely neutral. We take the
I\^{t}o's formula \cite{czuppon2021,wio1989path}:
\begin{eqnarray}\label{ito}
\Delta x=x_{t_n}-x_{t_{n-1}}=\mu(x_{t_{n-1}}) \epsilon +\sqrt{\sigma^2(x_{t_{n-1}})} (W_{t_n}-W_{t_{n-1}}),
\end{eqnarray}
where $W_{t}$ is the Wiener process (or Brownian motion)
\begin{eqnarray}
P(W_{t_2},\,t_2|W_{t_1},\,t_1)=\frac{1}{\sqrt{2\pi(t_2-t_1)\sigma^2(x_{t_1}) }}\exp\left [-\frac{1}{2(t_2-t_1)\sigma^2(x_{t_1}) }(W_{t_2}-W_{t_1})^2\right ]
\end{eqnarray}

For one-step update through the Moran model, we have
\begin{eqnarray}
\mu(x)&=&\pi^+-\pi^-,\nonumber\\[6pt]
\sigma^2(x)&=&\frac{\pi^+-\pi^-}{2N}.
\end{eqnarray}
Substituting Eq.~(\ref{pi}) into above equation and considering the weak selection (i.e. $\omega<<1$),
one has the perturbation expansion
\begin{eqnarray}
\mu(x)&\simeq &\omega x(1-x)\alpha (x-x^*),\nonumber\\[6pt]
\sigma^2(x)& \simeq &2 x(1-x),
\end{eqnarray}
with $\alpha=a-b-c+d$.

The transition probability would be \cite{wio1989path}
\begin{align}
p(x^{'},t^{'}|x_0,t_0)& =\int_{0}^{1}\int_{0}^{1}\cdots \int_{0}^{1} \text{d}W_1\text{d}W_{2}\cdots \text{d}W_{n-1} \left (\prod_j\frac{1}{\sqrt{ 4\pi\sigma^2(x_{t_{j}})  \epsilon} }\right )\times\nonumber\\[4pt]
 &\qquad\qquad \times \exp{  \left [\sum_j-\frac{1}{4\epsilon\sigma^2(x_{t_j}) }(W_{j+1}-W_j)^2\right ] }\nonumber\\[6pt]
&=\int_{0}^{1}\int_{0}^{1}\cdots \int_{0}^{1}\text{d}x_1\text{d}x_{2}\cdots \text{d}x_{n-1}  \left (\prod_j\frac{1}{\sqrt{ 4\pi\sigma^2(x_{t_{j}})  \epsilon} }\right ) \det\left (\frac{\partial W_j}{\partial x_k}\right )  \times\nonumber\\[4pt]
&\qquad\qquad\times \exp{  \left (\sum_j-\frac{1}{4\epsilon\sigma^2(x_{t_j}) }\left [(x_{j+1}-x_{j})- \mu(x_{t_{j}}) \epsilon \right ]^2\right ) },
\end{align}
where the Jacobian matrix is given from Eq.~(\ref{ito})
\begin{align}
\det\left (\frac{\partial W_{t_j}}{\partial x_k}\right ) &=\prod_{j=1}^N\frac{1}{\sqrt{\sigma^2(x_{t_{j-1}})}}.
\end{align}

As Schraiber method, we compute the relative probability density function for a path with selection compared to a neutral path,
\begin{align}\label{ps1}
p_s&=p_\omega(x^{'},t^{'}|x_0,t_0)/p_0(x^{'},t^{'}|x_0,t_0)\nonumber\\[3pt]
&=\int_{0}^{1}\int_{0}^{1}\cdots \int_{0}^{1} \text{d}x_1\text{d}x_{2}\cdots \text{d}x_{n-1}\times\nonumber\\[4pt]
&\qquad\times \exp{\left (\sum_j-\frac{1}{4\epsilon\sigma^2(x_{j}) }\left [(x_{j+1}-x_{j})- \mu(x_{t_{j}}) \epsilon \right ]^2 - \sum_j-\frac{1}{4\epsilon\sigma^2(x_j) }(x_{j+1}-x_{j}) ^2\right ) }\nonumber\\[4pt]
&=\int_{0}^{1}\int_{0}^{1}\cdots \int_{0}^{1} \text{d}x_1\text{d}x_{2}\cdots \text{d}x_{n-1}\times\nonumber\\[4pt]
&\qquad\times \exp{\left [-\sum_j\left(\frac{1}{4}\omega\alpha(x_{j+1}-x_j)(x_j-x^*) + \frac{1}{8}\omega^2\alpha^2x_j(1-x_j)(x_j-x^*)^2\epsilon \right)\right ] }\nonumber\\[4pt]
&=\int_{0}^{1}\int_{0}^{1}\cdots \int_{0}^{1} \text{d}x_1\text{d}x_{2}\cdots \text{d}x_{n-1}\exp{\left [-\sum_j \frac{1}{4}\omega\alpha(x_{j+1}-x_j)(x_j-x^*)\right ]}\nonumber\\[4pt]
&=\int  \mathscr {D}[x] e^{-\int ^t_{t_0}  \mathrm{d}t\frac{1}{4}\omega\alpha\dot{x}(x-x^*)   }=\int  \mathscr {D}[x] e^{-\int ^t_{t_0}  \mathrm{d}t \mathcal{L}[x,\,\dot{x}] },
\end{align}
where $p_{w(0)}(x_1,t_{1}|x_0,t_0)$ is the propagator from $x_0$ to $x_1$ when there is natural selection(and not), and
\begin{align}
\mathcal{L}[x,\,\dot{x}]=\frac{1}{4}\omega\alpha\dot{x}(x-x^*),
\end{align}
would be the stochastic Lagrangian for the Moran process in the two strategy evolutionary games. We can also identify the stochastic action of this process through
\begin{align}
\mathcal{S}[x(t)]=-\int ^t_{t_0}  \mathrm{d}t \mathcal{L}[x,\,\dot{x}].
\end{align}

\section{Discussion \& Conclusion}
In the present work, we developed the path integral formula for the Moran process of the stochastic evolutionary game theory in the finite population. Introducing the path integral approach, we here concentrate our attention on the transition probability, which
is the probability of the transition from the initial state to the final state. And it characterizes the evolution over time. It's no doubt that the fixation probabilities and the fixation time of an strategy, and the stationary distribution of the strategy
\cite{1979Mathematical,huang2010fixation} also can be explored to portray the evolutionary dynamics of the strategy frequency. However, the transition probability actually has some special applications. For example, there recently has been growing interest in
analyzing samples taken from the same or related populations at different time points by the transition probability \cite{hummel2005detection,bollback2008estimation,green2010draft,reich2010genetic}. Hence, it is important to calculate the transition probability.

Making use of the transition probability, the probability density (i.e., the frequency of the strategy in our work) at a certain time can be obtained. It has been calculated by some other methods in the previous works
\cite{tran2013introduction,chalub2014frequency,burden2019stationary}. For the stochastic process without including the selection intensity, the probability density has been calculated by the solution which is obtained based on the Fokker-Plank equation. However,
for the situation that the selection intensity is considered, it's invalid for Fokker-Planck equations to get an analytical approximate solution \cite{yu1997new}. Based on the formulation of the transition probability, it's possible to obtain the probability
density at a certain time given the probability density at the initial time. Specifically, the probability density $p(x,t)$ of the evolution $X(t)$ is then obtained from
\begin{equation}
p(x,t)=\int_{0}^{1}p(x,\,t|x_0,t_0)p(x_0,t_0)\mathrm{d}x_0,
\end{equation}
where $p(x_0,t_0)$ is the initial probability density of $x(t)$ at $t=t_0$. According to the path integral formula we mentioned, we obtain
\begin{eqnarray}
p(x,t)&=&p(x_0,t_0) \int_{R}p(x,\,t|x_{n-1},t_{n-1})p(x_{n-1},t_{n-1}|x_{n-2},t_{n-2})\times \cdots\nonumber\\[12pt] && \qquad \qquad \qquad \cdots \times p(x_2,t_2|x_1,t_1)p(x_{1},t_{1}|x_0,t_{0})\mathscr{D}x,
\end{eqnarray}
where $R$ is the range of the \emph{n}-dimensional state for $x$. With the help of the transition probability, on the one hand, the probability density at any time during the evolutionary process, not just the stationary distribution, can be calculate. On the
other hand, the probability density for the cases including the selection intensity, or other factors we will illustrate bellow, also could be calculated.

As a new method to calculating the transition probability, the path integral can be applied not only when the selection intensity is constant, but also when it is frequency-dependent. We explored the relative probability density function for a path as Eq.~(\ref{ps1}), in which the integral with respect to a neutral path can be obtained directly. The neutral intermediate evolutionary process can be regarded as the normalized coefficient. The more exact model to reflect a realistic scenario of natural evolution is
that the selection intensity is not constant. For example, it may be frequency-dependent selection, which has been widely studied in the framework of evolutionary game \cite{ayala1974frequency,alexander2002evolutionary,imhof2006evolutionary,traulsen2007pairwise}
and eco-evolutionary game \cite{gokhale2016eco,czuppon2018disentangling,govaert2019eco}. Consequently, it's reasonable to change the constant selection intensity as the function of the frequency, i.e. $w(x)$. The propagator thus is the function of the system
state.

In a word, the path integral approach provides a novel insight to study the evolutionary dynamics. We will consider the evolutionary process with eco-evolutionary feedback, environment fluctuation, the distinguishing selection intensity \cite{kandori1993learning, sample2017limits,wang2019evolutionary}, mutation \cite{sniegowski1997evolution,taddei1997role,willensdorfer2005mutation} and
so on in this framework. And how these factors affect the evolutionary process could be explored in this formula in the future.

\section{ACKNOWLEDGEMENT}

This project was supported by the National Natural Science Foundation of China (No. 32171482).


\end{document}